\UseRawInputEncoding

\documentclass[conference]{IEEEtran}

\usepackage[dvips]{graphicx}

\ifCLASSINFOpdf
\else
\fi
\begin{document}
\title{Proposal of Automatic Offloading Method in Mixed Offloading Destination Environment}

\author{\IEEEauthorblockN{Yoji Yamato}
\IEEEauthorblockA{NTT Network Service Systems Laboratories\\
NTT Corporation\\
Musashino-shi, Tokyo 180--8585, Japan\\
Email: yoji.yamato.wa@hco.ntt.co.jp}}


%Telephone: (800) 555--1212, Fax: (888) 555--1212}


\maketitle

\begin{abstract}
When using heterogeneous hardware, barriers of technical skills such as OpenMP, CUDA and OpenCL are high. Based on that, I have proposed environment-adaptive software that enables automatic conversion, configuration. However, including existing technologies, there has been no research to properly and automatically offload the mixed offloading destination environment such as GPU, FPGA and many core CPU. In this paper, as a new element of environment-adaptive software, I study a method for offloading applications properly and automatically in the environment where the offloading destination is mixed with GPU, FPGA and many core CPU. 
\end{abstract}


\begin{IEEEkeywords}
Environment Adaptive Software, GPGPU, Automatic Offloading, Mixed Offloading Destination Environment.
\end{IEEEkeywords}

% \fi
\IEEEpeerreviewmaketitle

\section{Introduction}
It is said that Moore's Law will be slow down and a central processing unit's (CPU's) transistor density cannot be expected to double in 1.5 years. Based on this situation, systems with heterogeneous hardware, such as graphics processing units (GPUs), field-programmable gate arrays (FPGAs) and many core CPUs are increasing. For example, Microsoft's search engine Bing uses FPGAs \cite{8}, and Amazon Web Services (AWS) provides GPU and FPGA instances \cite{7} using cloud technologies (e.g., \cite{29}-\cite{17}). 

However, in order to properly utilize devices other than small core CPUs in the system, it is necessary to make configurations and programs that consider device characteristics (e.g., \cite{39}-\cite{16}), and knowledge of OpenMP (Open Multi-Processing) \cite{OpenMP}, OpenCL (Open Knowledge of Computing Language) \cite{12} and CUDA (Compute Unified Device Architecture) \cite{11} is required. Therefore, for most programmers, skill barriers are high. In addition, the use of IoT devices (e.g., \cite{1}-\cite{4}) and external service coordination (e.g., \cite{5}-\cite{10}) are increasing and required skills are also increasing.

The expectation of applications using heterogeneous hardware is getting higher; however, the skill hurdles are currently high for using them. To surmount such hurdles, it will be required that application programmers will only need to write logics to be processed, then software will adapt to the environments with heterogeneous hardware to make it easy to use such hardware.

Therefore, I previously proposed environment-adaptive software that effectively runs once-written applications by automatically executing code conversion and configurations so that GPUs, FPGAs, many core CPUs, and so on can be appropriately used on deployment environments \cite{GPU2}. As an element technology of environment-adaptive software concept, I also proposed a method for automatically offloading loop statements and function blocks of application to GPUs or FPGAs \cite{13}. 

The purpose of this paper is to automatically offload applications with high performances in mixed offloading destination environments in which various devices of GPU, FPGA and many core CPU exist. First, I propose a method that automatically offloads to a single device of GPU, FPGA or many core CPU. Next, I propose a method of appropriate offloading in mixed offloading destination environments with various devices. 
\section{Proposal of automatic offloading to mixed offloading destination environment with various devices}

Based on previous elemental technologies, this section describes the basic idea for various devices, offloading to each device and proposes the automatic offloading method to mixed offloading destination environment.

\subsection{Basic ideas for various devices offloading}

The various offloading destination environments cover in this paper are the GPU, FPGA and many-core CPU. GPUs and FPGAs have a long history as heterogeneous hardware different from small core CPUs, and there are many cases of speed-up by manual offloading using CUDA and OpenCL, and the market is large. Regarding many-core CPUs, CPUs equipped with a large number of cores of 16 cores or more have recently come on the market even at a low price of a few thousand dollars, and parallelization is performed using technical specifications such as OpenMP. There are many cases in which the speed is improved by manual tuning.

In order to perform automatic high-speed offload, I take the approach of gradually searching high-speed offload patterns with evolutionary calculation by measuring the performance on a physical machine in a verification environment. This is the same as in the case of GPU offload, which has been proposed so far. The reason for this is that the performance varies greatly not only with the code structure but also with the actual processing contents such as the specifications of the processing hardware, the data size, and the number of loops. Therefore, it is difficult to predict the performance statically and it is necessary to conduct dynamic measurements, I think. In the market, there is an automatic parallelizing compiler that finds loop statements and parallelizes them at the compile stage. However, it is often necessary to measure performance because parallelization of parallelizable loop statements often results in low speed.

As for the object to be offloaded, I take the approach of focusing the loop statement and function block of the program. This is the same as in the case of GPU, FPGA offload, which have been proposed so far. With regard to loop statements, loop statements are the first target for offloading because most of the processing of programs that take a long time is spent in loops. On the other hand, with regard to function blocks, when speeding up specific processing, an algorithm suitable for the processing content and processing hardware is often used, so there is a case where it can be greatly speeded up compared to offloading individual loop statements. Performance is improved by replacing frequently used functional blocks such as Fourier transform to the processing which are implemented based on appropriate algorithms suitable for devices characteristics.

\subsection{Automatic offloading to each offloading device}
I describe automatic offloading for three offloading destination of GPU, FPGA and many-core CPU by two methods, loop statement and function block offloading. 

\subsubsection{Automatic many core CPU offloading of loop statements}
I propose automatic many core CPU offloading method of loop statements, newly.

Like GPUs, many core CPUs utilize many computational cores and parallelize processing to speed up. Unlike the GPU, the many-core CPU has a common memory, so it is not necessary to consider the overhead due to data transfer between the CPU and the GPU memory, which is often a problem with offloading to the GPU. In addition, the OpenMP specification is frequently used for parallelizing program processing on a many core CPU. OpenMP is a specification that specifies parallel processing and other processing for a program with directives such as \#pragma omp parallel for. The OpenMP programmer is responsible for parallelizing the processing in OpenMP. When the processing that cannot be parallelized is parallelized, the compiler does not output an error and the calculation result becomes wrong.

Based on these, in this paper, I take the approach of genetic algorithm method \cite{20} that gradually accelerates the offload patterns for many core CPUs automatic offloading of loop statements. Specifically, multiple patterns are created that specify whether parallel processing of loops is conducted or not with OpenMP \#pragma directives, and repeats actual performance measurement in the verification environment. Here, PGI compiler used for automatic GPU offload output an error when the parallelization was impossible. However, OpenMP compilers such as gcc do not output such errors. Therefore, in order to automate, the processing performed by the OpenMP directive is simplified to only whether the loop statement is processed in parallel by the many-core CPU, only the pattern that produces the correct calculation result remains in the evolutionary calculation by checking whether or not the final calculation result is correct for each measurement.

The proposed method behaves as follows (Figure 1). When the code is input, the syntax is analyzed and the loop statement is determined. By adding \#pragma omp parallel for to the loop statement, OpenMP code that specifies parallel processing is created. Here, the gene pattern is set to 1 when parallel processing is performed on the many core CPU and 0 when parallel processing is not performed. The proposed method compiles the prepared multiple patterns with an OpenMP compiler such as gcc, and measures the performance in a verification environment machine equipped with a many core CPU. As a result of performance measurement, the pattern of high-speed processing is set to high goodness of fit and the pattern of low-speed processing is set to low goodness of fit, and next-generation patterns are created by processing such as elite selection, crossover and mutation of the genetic algorithm. Here, in the performance measurement, the fact that the final calculation result is the same as that without parallel processing is compared with the case where the original code is processed with a single core of CPU. If the difference is unacceptably large, the goodness of fit of the pattern is set to sufficiently low so that it will not be selected for the next generation.

 \begin{figure}[tb]
 \begin{center}
  \includegraphics[width=85mm]{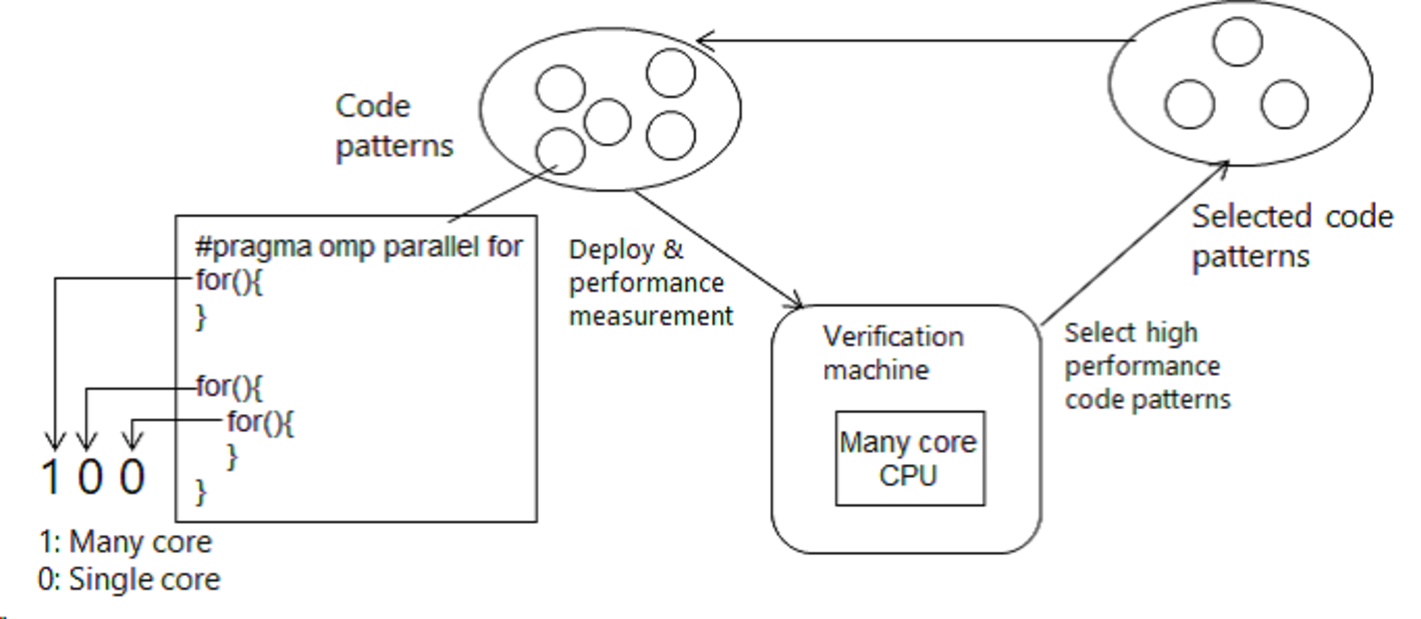}
 \end{center}
 \caption{Automatic many core CPU offloading method of loop statements}
   \end{figure}

\subsubsection{Automatic GPU offloading of loop statements}
I proposed \cite{GPU2}\cite{GPU3} for automatic GPU offloading method of loop statements. Automatic offload is enabled by appropriate loop statements extraction using genetic algorithm and reduction of CPU-GPU memory data transfer.

\subsubsection{Automatic FPGA offloading of loop statements}
I proposed \cite{FPGA} for automatic FPGA offloading method of loop statements. The candidate loop statements are narrowed down by using the arithmetic intensity, the number of loops, and the amount of resources. After narrowing down, the performance of multiple patterns in the verification environment is measured to enable automatic offloading.

\subsubsection{Automatic offloading of function blocks}
I proposed \cite{Block} for automatic offloading method of function blocks. The method was verified for GPU and FPGA, but the processing is common to detect function blocks that can be offloaded by name matching and similarity detection, so it can be adopted in the many core CPU as well.

\subsection{Automatic offloading for mixed offloading destination environment}
Regarding performance, even if the program code is the same, it depends on the specifications of the processing hardware and the processing contents (data size, number of loops, etc). Therefore, actual performance measurements are needed. Since there are three types of offloading devices; many core CPU, GPU and FPGA and there are two methods of offloading focus points; loop statement and function block, six (3*2) offload verifications are needed at least.

Comparing to loop statements and function blocks, function block offloading can be faster if there are offloadable functions. Comparing to prices of many core CPU, GPU and FPGA, the central price range is the ascending order of GPU, many core CPU and FPGA. Comparing to verification time for one pattern, verification time is the ascending order of many core CPU, GPU and FPGA. Currently, FPGA requires several hours for circuit implementations, and it needs a time to measure performance.

Based on these situations, I propose the order of verification with six offloads is as follows; many core CPU function block offload, GPU function block offload, FPGA function block offload, many core CPU loop statement offload, GPU loop statement offload, FPGA loop sentence offload. Offload verifications are performed in this order to search for high-performance patterns. The target code of the first three and the latter three may be different. Specifically, if functional block offloading is possible in the first three, the latter three loop statement offloading is verified on the code without the function block part that was offloadable. In the offload verifications, the user can specify the target performance and price, and if a sufficiently fast and low-priced offload pattern is found in front of the six verifications within the range specified by the user, the subsequent verifications will not be performed. 

The reason for this is that function block offloading can be faster, so verification will be performed ahead. Also, when performing automatic offloading, it is expected that high-speed pattern search will be performed at the low cost and in a short time as possible. Therefore, the FPGA that requires long verification time is the last. In addition, FPGA verification is not performed if a pattern that satisfies the user requirements is found in the previous stage. Regarding many core CPU and GPU, there is no big difference in terms of price and verification time, many core CPU has a smaller difference from small core CPU than GPU which has a separate memory space. Therefore, the verification order follows above description. If a pattern that sufficiently satisfies the user requirements is found in the many core CPU ahead, GPU verification will not be performed.

\section{Evaluation}
Since the automatic offload of loop statements to GPU and FPGA are evaluated by \cite{GPU2}\cite{FPGA}, and the automatic offload of function blocks to GPU and FPGA is evaluated by \cite{Block}, this paper confirms that applications can be offloaded to appropriate devices in mixed offloading destination environment. 

\subsection{Evaluation method}
\subsubsection{Evaluated applications}
Based on the implementation, I evaluate applications of matrix calculation, block tridiagonal solver and signal processing. 

Simple matrix calculation is used in many types of analysis such as machine-learning analysis. Because matrix calculation is used not only on cloud sides but also device sides due to the spread of IoT and AI, automatic performance improvements for various applications are needed. For experiment, I use polybench 3mm (3 matrix multiplications) in which three matrix multiplications are calculated with a size of 1000*1000 (STANDARD\_DATASET: NI=1000, NJ=1000, NK=1000, NL=1000, NM=1000) \cite{3mm}.

The block tridiagonal solver is a numerical solution of partial differential equations. There are various kinds and implementations of numerical calculation, I use NAS.BT (NASA Block Tridiagonal Solver) \cite{NAS} as an example of medium size numerical calculation application with more than 100 loop statements. The parameters are CLASS A settings with grid size=64*64*64, number of iterations=200 and time step=0.0008.

Among signal processing, the time-domain finite-impulse response filter performs processing in a finite time on the output when an impulse function is input to a system. When considering applications that transfer signal data from devices over the network, to reduce network costs, it is assumed that signal processing such as filters are conducted on devices sides, thus signal processing offloading is important. I use MIT Lincoln laboratory's high performance embedded computing (HPEC) Challenge Benchmark C code of tdFIR (time-domain finite-impulse response filter) \cite{FIR}. HPEC's sample test processing is carried out with 64 filters and 4096 length of input/output vectors.

\subsubsection{Experiment conditions}
In this experiment, the implementation receives codes, it parses by Clang \cite{21} and verifies offloading of function block and loop statements for three offloading destinations of GPU, FPGA and many core CPU. Based on six verifications, the implementation selects or creates a high performance pattern and measures the performance. I evaluate the degree of improvement compared to the case where all processing is done by a single core of CPU without offloading. At the same time, I confirm that the appropriate device is selected.

The experimental conditions are as follows.

Offload applications and loop statements number: Matrix multiplications 3mm which has 20 loops, Block tridiagonal solver NAS.BT which has 179 loops, Time-domain finite-impulse response filter tdFIR which has 6 loops.

\medskip
The experimental conditions of function block offload are as follows.

Offload targets: Intel sample OpenCL of time-domain finite-impulse response filter \cite{Intelsample}. In this evaluation, I prepare one function block offload target because I only need to confirm appropriate device and method selection.
 
Offloadable function block discovery method: The code of the offload source application tdFIR calls the corresponding method on the code side and discovered by DB name matching and also by Deckard \cite{44}. 

\medskip
The experimental conditions of the GA for GPU and many core CPU loop statement offload are as follows.

Gene length: Number of GPU and many core CPU processable loop statements. 18 for 3mm, 120 for NAS.BT and 6 for tdFIR.

Number of individuals M: No more than the gene length. 16 for 3mm, 20 for NAS.BT and 6 for tdFIR.

Number of generations T: No more than the gene length. 16 for 3mm, 20 for NAS.BT and 6 for tdFIR.

Goodness of fit: (Processing time)$^{-1/2}$. When processing time becomes shorter, the goodness of fit becomes larger. By setting the power of (-1/2), I prevent the narrowing of the search range due to too high of the goodness of fit of specific individuals with short processing times. If the performance measurement does not complete in 3 minutes, a timeout is issued, and processing time is set to 1000 seconds to calculate goodness of fit. If the calculation result is largely different from original codes result, the processing time is also set to 1000.

Selection algorithm: Roulette selection and Elite selection. Elite selection means that one gene with maximum goodness of fit must be reserved for the next generation without crossover or mutation.

Crossover rate Pc: 0.9

Mutation rate Pm: 0.05

\medskip

The experimental conditions of FPGA loop statements offload were as follows.

Narrowing down using arithmetic intensity: Narrow down to the five loop statements by arithmetic intensity and loop count with ROSE \cite{ROSE} and gcov \cite{gcov}.

Narrowing down using resource efficiency: Narrow down to the top three loop statements in resource efficiency analysis. The implementation selects the top three loop statements with high arithmetic intensity/resource amount in this verification.

Number of measured offload patterns: 4. In the first measurement, the top three loop statement offload patterns were measured. Then, the second measurement was measured with the combination pattern of two loop statement offloads that were high performance at the first measurement.

\subsubsection{Experimental environment}
I use AMD Ryzen Threadripper 2990WX as a many core CPU device. Gcc 10.1 processes OpenMP for many core CPU control. I use GeForce RTX 2080 Ti as a GPU device. PGI compiler 19.10 and CUDA toolkit 10.1 process OpenACC for GPU control. Ryzen CPU and GeForce GPU are equipped on the same node. I use Intel Arria 10 GX FPGA as a FPGA device. Intel Acceleration Stack 1.2 process OpenCL to control FPGA which is installed on DELL EMC PowerEdge R740. Figure 2 shows the experimental environment and environment specifications. A client note PC specifies the C/C++ application codes, which are converted and vivificated on verification machines. 

 \begin{figure}[tb]
 \begin{center}
  \includegraphics[width=85mm]{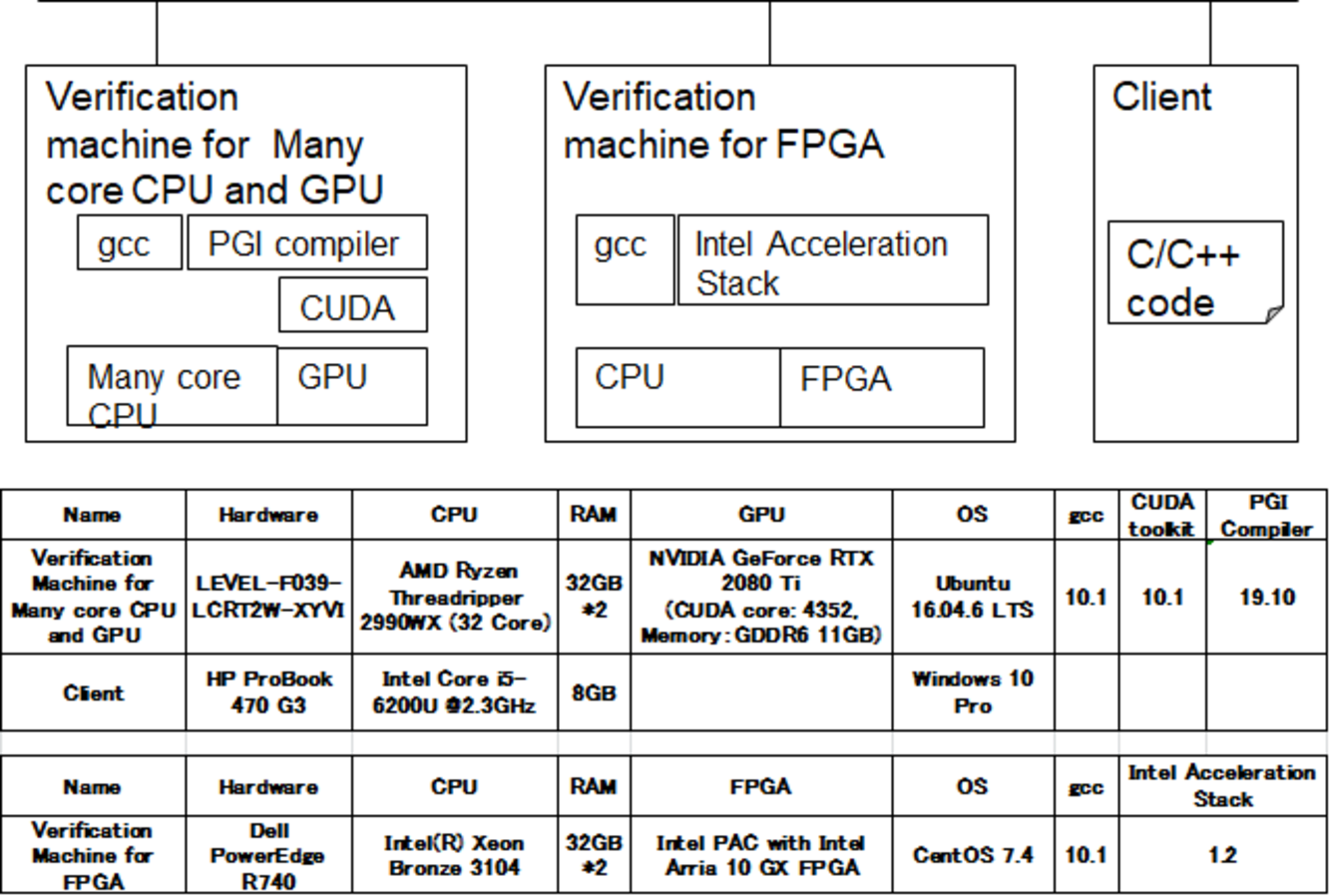}
 \end{center}
 \caption{Experimental environment}
   \end{figure}

\subsection{Performance results}
I confirmed the high performances with various devices.

For 3 applications, Figure 4 shows the processing time by a single core of CPU, which device and which method was used for offloading, the offload processing time, the degree of performance improvement, and the results of offloading to another device. The degree of performance improvement shows how much improve processing time of applications. 1 means the same processing time when a sample test is processed by a single core of CPU on the offload destination machine after gcc or PGI compiler compile. Regarding 3mm, the processing time with a single core was 51.3 seconds, but by GPU with loop statement offload, it was processed in 0.046 seconds, and the improvement was 1120 times. When offloading loop statements to many core CPU, it was processed in 1.05 seconds and the improvement was 44.5 times, and a higher performance GPU offloading was selected. Regarding NAS.BT, the processing time with a single core was 130 seconds, but by many core CPU with loop statement offload, it processed in 24.1 seconds, and the improvement was 5.39 times. When the implementation tried to offload loop statements to GPU, the processing time exceeded 130 seconds, so the result was 130 seconds processed by a single core of CPU without any offload. Regarding tdFIR, the processing time in a single core was 0.298 seconds, but by FPGA with function block offload, it processed in 0.0142 seconds, and the improvement was 21.0 times. Since the function block offloading was possible, the loop statement offloading of the remained part where the function block was removed was not attempted. However, if the function block offloading was not performed and the looping statement offloading was performed by FPGA, the improvement was only 4.00 times.

For 3 of the function block offloads in the 6 measurements, the search only takes about few minutes, but when using FPGA, it takes about 3 hours to implement the replaced codes. For 3 of the loop statement offloads in the 6 measurements, FPGA verifications take half a day because it takes 3 hours for 1 pattern, and searching by genetic algorithm with many core CPU or GPU takes about 6 hours each. As a result, all measurements take about a day.

 \begin{figure}[tb]
 \begin{center}
  \includegraphics[width=85mm]{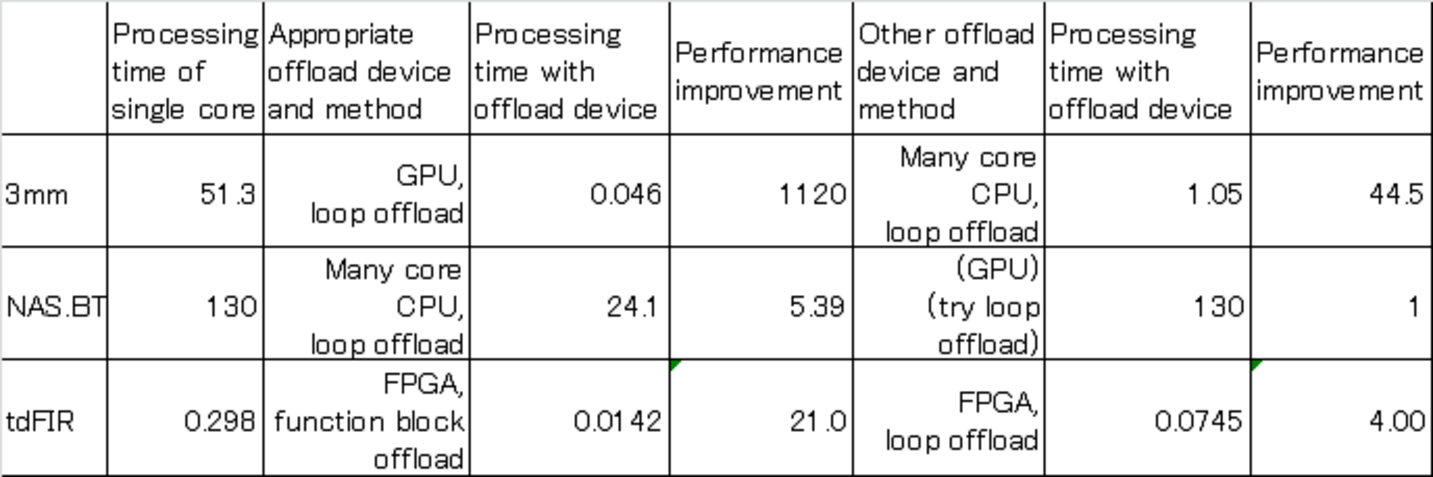}
 \end{center}
 \caption{Results of offloading to mixed offloading destination environment}
   \end{figure}

\section{Conclusion}
I proposed an automatic offloading method for mixed offloading destination environments with various devices of GPU, FPGA and many core CPU as a new element of my environment-adaptive software.

First, as a preparation, I proposed an automatic offload method for loop statements for a many core CPU as one of various offloading destination environments, with reference to the evolutionary computation method for a GPU. Next, I studied the order of offload trials for each offloading device and the speedup method when there were multiple offload candidates with one node. Specifically, the function block offload for many core CPU, the function block offload for GPU, the function block offload for FPGA, the loop statement offload for many core CPU, the loop statement offload for GPU, the loop statement offload for FPGA are verified in this order. This is because functional blocks offloading can be faster than loop statements offloading, and FPGA verifications take longer time to measure performance. 

I implemented the proposed method, evaluated its automatic offloading of several applications to mixed offloading destination environments, and confirmed its effectiveness.

\end{document}